# Profinet vs Profibus

Pouya Aminaie[1] and Poorya Aminaie[2]

[1]Department of ECE, Shiraz University, Shiraz, Iran
[2] Department of ECE, Shahid Beheshti University, Tehran, Iran
E-mail: [2]p.aminaie@mail.sbu.ac.ir

**Abstract**

We present a step by step definition of Profinet and Profibus. We introduced different types of each of the two communication protocols. We then described the topology and performance of each one individually. Finally, the properties of them have been compared to show that which one has better performance in the industry.

**1. Introduction**

1.1.   Profinet

Profinet is the abbreviation for Process Field Net, which refers to technical standards for data communication through Ethernet in the industry. These types of standards are used for gathering data and controlling industrial equipment. As can be seen from Fig.1, Profinet satisfies all the needs of industrial technologies.

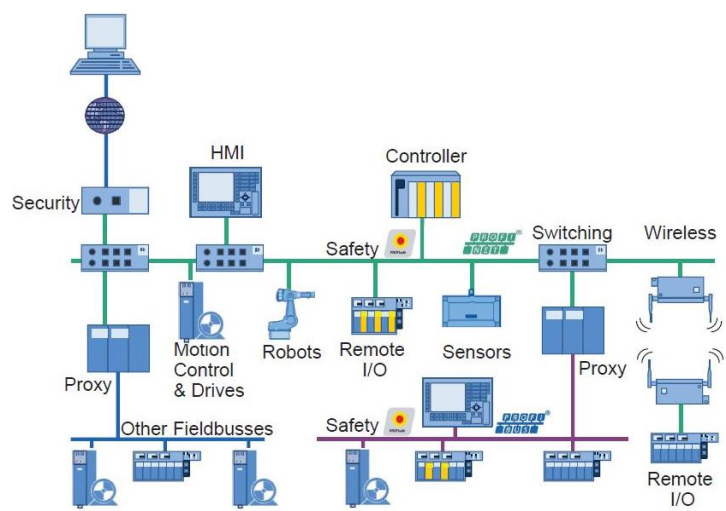

Fig.1 Requirement of automation technology [1]

The need for Profinet is felt in production automation and processing automation sections, where its use can resolve many of these needs. Profinet can be divided into two main categories, as follows:

- Profinet IO
- Profinet CBA

1.2. Profibus

The word Profibus is taken from the phrase Process Field Bus. The scope of this protocol covers from the field level to the control level. The advantages of Profibus are as follows:

1. Low noise acceptance due to twisted pair cable being the transmission interface.
2. Appropriate bandwidth due to the use of an appropriate transmission method such as RS485.
3. Secure and non-interfering data exchange for using the token pass access method.
4. High network flexibility due to the open-source nature of this protocol.

This protocol uses the ISO standard within its layers. However, this standard does not cover all the layers. As shown in Fig.2, layer 1, 2 and 7 are only used.

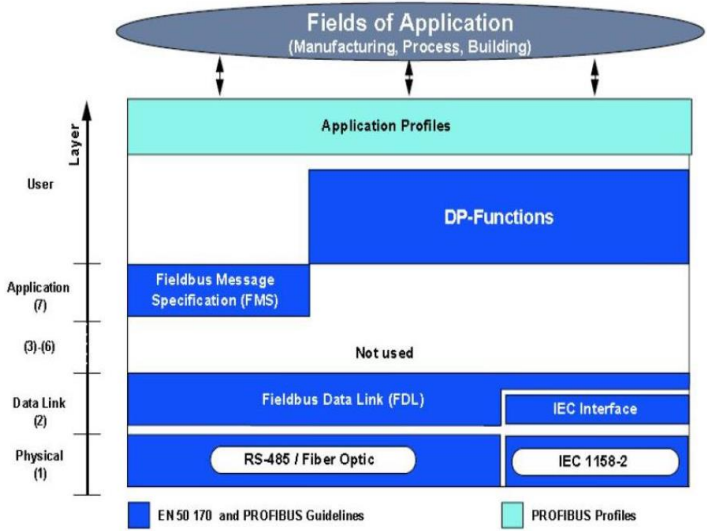

Fig.2 Profibus layers in use

Profibus can be divided into three main categories, as follows:

- Profibus DP
- Profibus FMS
- Profibus PA

In the following, we will introduce each of the two types of protocols and the structure of each type.

**2. Profinet Structure**

2.1. Categories:

  2.1.1. Profinet IO

This type of Profinet is used for large scale applications in a modular format, and the devices used in it include the following three parts:

- IO-Controller: Controls the automation tasks.
- IO-Device: is a field device, controlled and monitored by IO-Controller.
- IO-Supervisor: is generally used for recognizing the IO-Device and is used as network software.

  2.1.2. Profinet Component Based Automation (CBA)

This type of Profinet is also used for large scale applications and is implemented modularly, with the difference that it is more intelligent compared to Profinet IO. The relationship between the two devices is a machine-to-machine relation.

  2.1.3. Differences Between Profinet IO and Profinet CBA:

These two types of Profinet deal with industrial Ethernet-based controllers from two different points of view, the differences of which are presented in Fig.3.

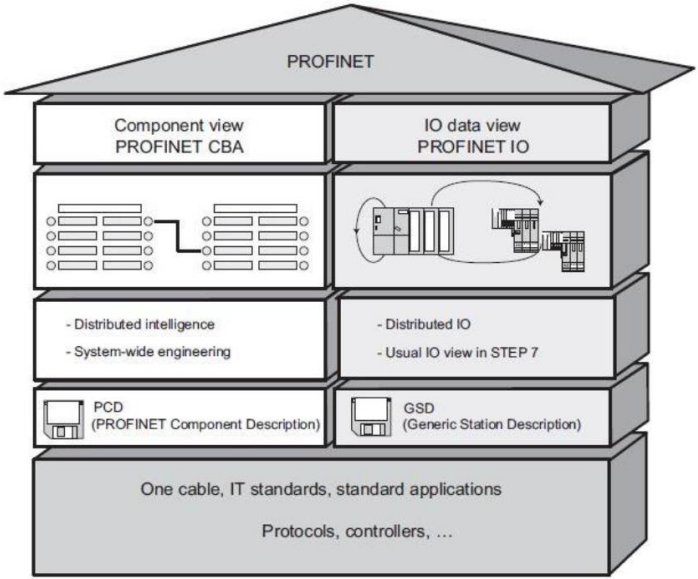

Fig.3 Difference between profinet IO and profinet CBA [2]

Profinet CBA divides the whole system into different functions that are configured and programmed. However, in Profinet IO, each of these controllers is programmed separately.

2.4. Physical Layering

- Electric cables: such as the use of a copper wire pair.
- Optical fiber cables: such as optical fibers are used in mono-mode or multi-mode formations. In the first case, the maximum distance between two nodes could be 26 kilometers, while in the second case, the distance between the two nodes cannot exceed 3000 meters.

An overview of Profinet layers is shown in Fig.4.

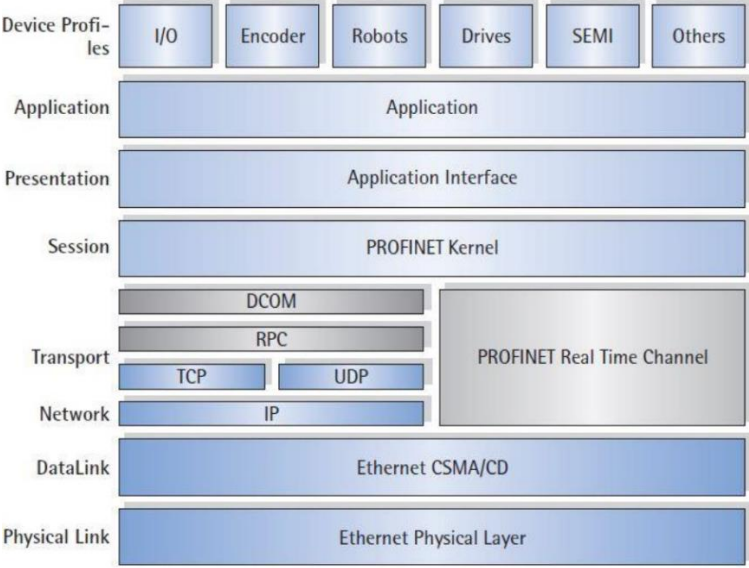

Fig.4 Profinet layers [3]

2.5. Protocols used in Profinet

- TCP/IP protocol: is used in processing automation. It is used to process not for essential data. Thus, its performance time is about 100ms.
- Real-time (RT) protocol: is used for Profinet IO, whose application is in factory automation. Its performance time is about 10ms.
- Isochronous Real-Time (IRT) Protocol: is used to control and drive the systems in Profinet IO. The performance time of this protocol is less than 1ms.

Fig.5 provides an overview of these three protocols.

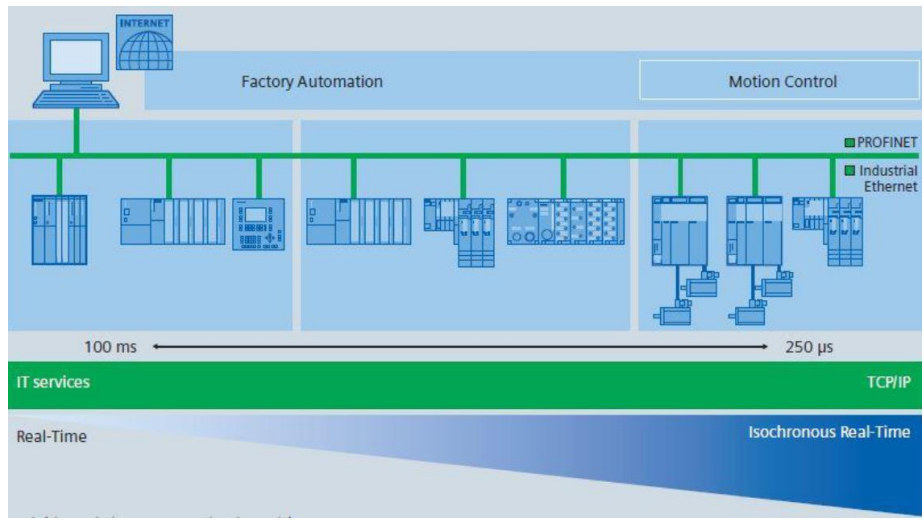

Fig.5 An overview of profinet protocols [4]

2.6. Addressing

Three types of addresses must be defined in Profinet as follows:

- MAC address
- IP: must be unique throughout the network.
- Device name

Two different protocols, namely discovery and configuration protocol (DCP) and dynamic host configuration protocol (DHCP), are required to assign these three addresses to the nodes. Each of these protocols is separately explained in detail in the following.

2.7. DHCP protocol

It is a protocol used by network devices to obtain various parameters required for the performance of resource-intensive programs in the IP. By using this protocol, the workload of system management is significantly reduced, and devices can be added to the network with a minimum amount of manual adjustments or no adjustments at all.

The DHCP performance could be divided into four essential functions:

- DHCP Discovery
- DHCP Offer
- DHCP Request
- DHCP Acknowledgement

These four stages are known as DORA for short, where each letter is taken from the beginning of each stage's name.

The DHCP messages structure is carried in UDP datagram, which uses port 67 on the server-side and port number 68 on the client-side. The general structure of the DHCP protocol is presented in the following table.

| OP | HTYPE | HLEN | HOPS |
|---|---|---|---|
| TRANSACTION ID ||||
| SECS || FLAGS ||
| CIADDR (Client IP address) ||||
| YIADDR (Your IP address) ||||
| SIADDR (Server IP address) ||||
| GIADDR (Gateway IP address) ||||
| CHADDR (Client hardware address (16 OCTETS)) ||||
| SERVER HOST NAME (64 OCTETS) ||||
| BOOT FILE NAME (128 OCTETS) ||||
| OPTIONS (VARIABLE) ||||

Fig.6 DHCP protocol structure [2]

2.8. DCP protocol

This protocol is based on the layering of data links used for addressing Profinet nodes.

2.9. Switching Technology

A bit rate of about 100mb/sec is used for switching in Profinet. The connection between every two nodes is a half-duplex two-sided connection, i.e., each node can act both as a transmitter and as a receiver.

2.10. The topology of the network

Different topologies are used to implement Profinet, four of which are introduced below. Fig.8 provides an example of a hybrid topology.

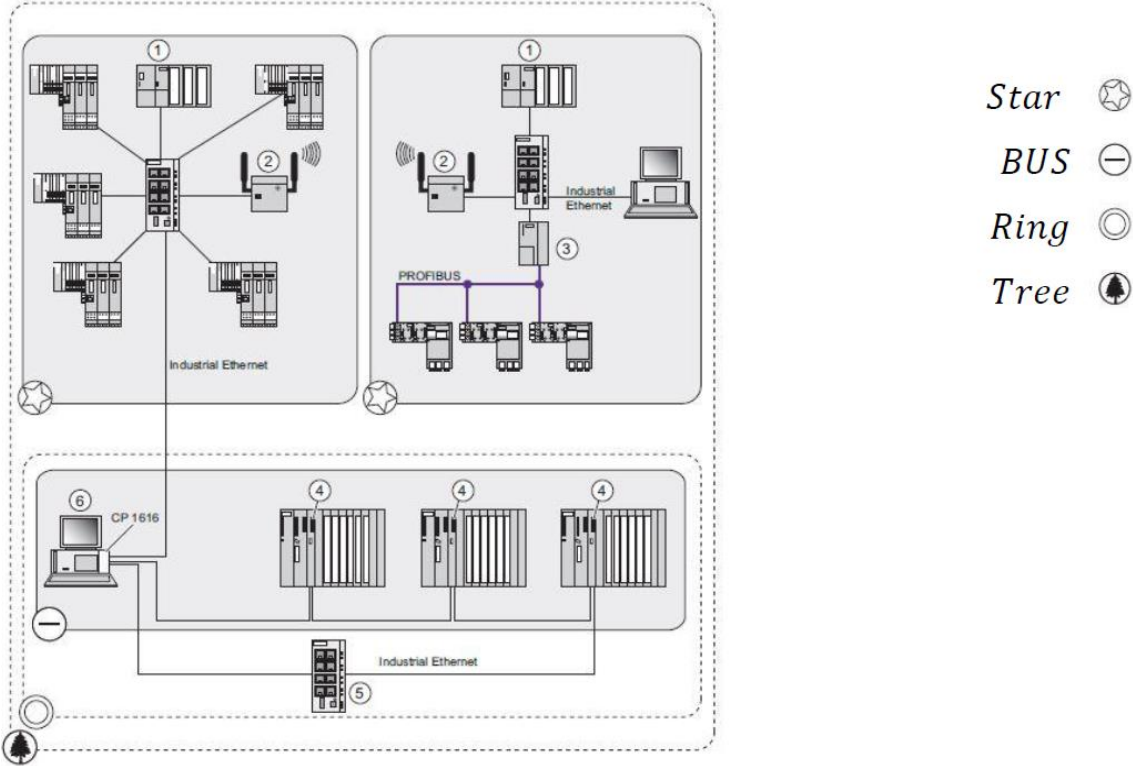

Fig.8 Hybrid Topology [2]

## 3. Profibus Structure

3.1. Profibus DP

It uses layers 1 and 2, along with a user interface. Layers 3 and 7 are not utilized in this protocol. This structure also enables fast data transmission. Its advantage over FMS is that the seventh layer is eliminated, which results in the optimization of the performance speed.

DP works in the form of a master/slave. In other words, the central core, or master, cyclically reads the inputs from the slaves and sends them back the outputs. The DP structure can transmit 512 bites of output data and 512 bites of input data over 32 slaves at each millisecond, with the speed of 12Mbps.

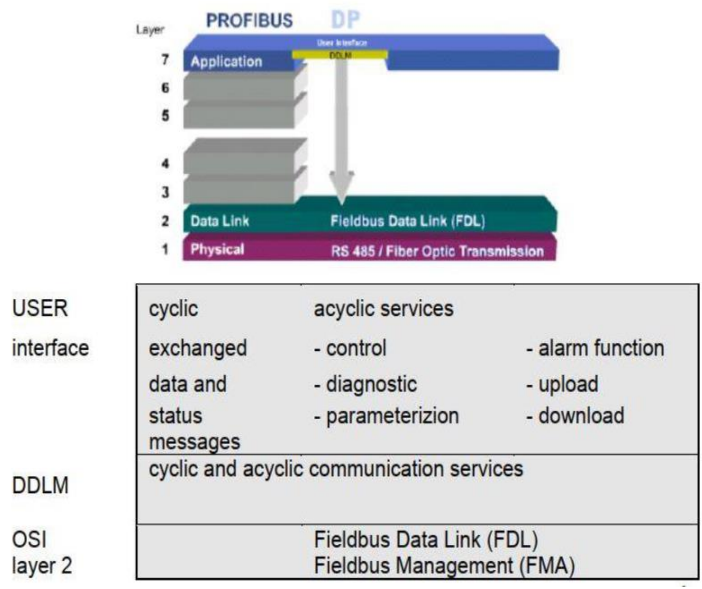

Fig.9 DP structure

In DP-V0, only cyclical connections between the master and the slaves are allowed, i.e., the master calls the slaves cyclically, one after another, and exchanges data with them. When a master wants to communicate with a slave, first the following initialization steps should take place:

1. The master sends a diagnostic request, and the slave reports its status.

2. Master sends the slaves parameters required for data transmission, such as response time. Moreover, the slave acknowledges them to the master.

3. The master sends the configured software structure to the slave. Moreover, the slave notifies the master if it recognizes any differences between this and the main structure.

The ability of non-cyclic connection is also added in DP-V1. Generally, two types of masters are defined in the DP protocol:

1. DP Master class 1, or DPM1, which is the same central core establishing cyclical connections.

2. DP Master class 2, or DPM2, which is connected to the slave during the setup or fault detection or for configuration, calibration. Its connection is non-cyclical and temporary, and it is not required to be permanently connected to the bus.

Therefore, when DPM1 reaches the last slave, it will assign the remainder of the work to DPM2 to conduct an acyclic connection to any slave required in the remainder of the bus cycle time, and then the routine is again assigned back to DPM1.

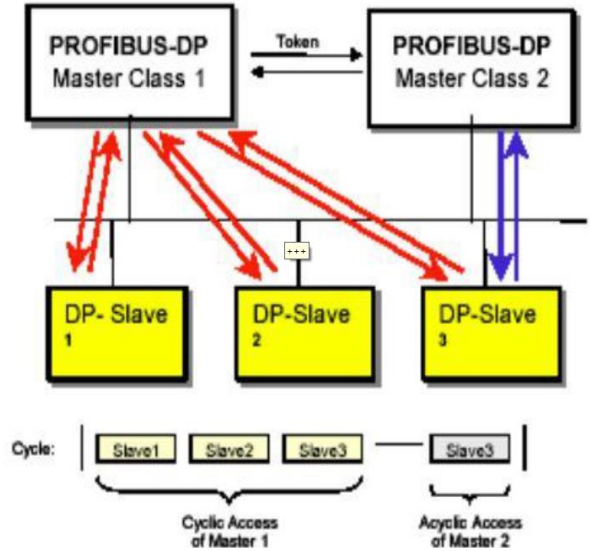

Fig.10 DP-V1 master-slave communications

In DP-V2, the capability of direct data transmission among the slaves is also provided. This method is in the form of a broadcast, i.e., one slave puts its data on the bus as a publisher for other slaves or subscribers to use.

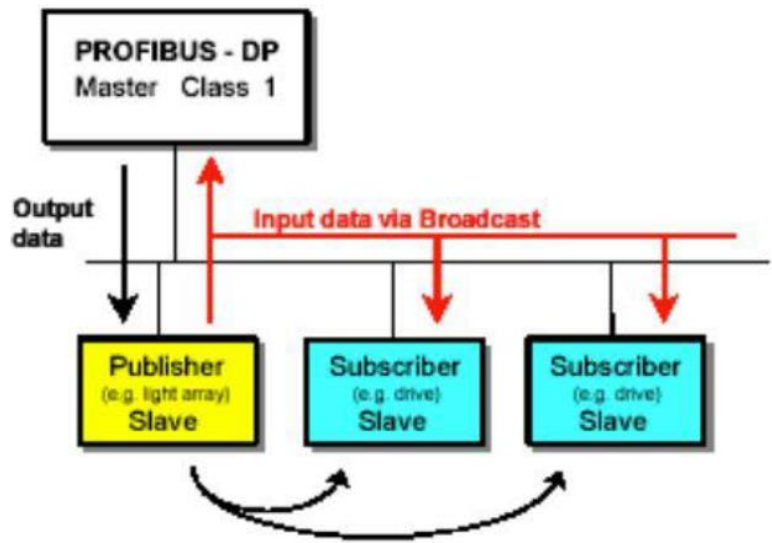

Fig.10 DP-V2 slave-slave communications

3.2. Protective mechanisms

    1. There is a separate timer for each slave in the central core, which sends a stop command if correct data are not received from the slaves during a specified time.

    2. There is a watchdog for each slave, which reveals the master or data transmission errors. Fig.11 shows this two prospective mechanism.

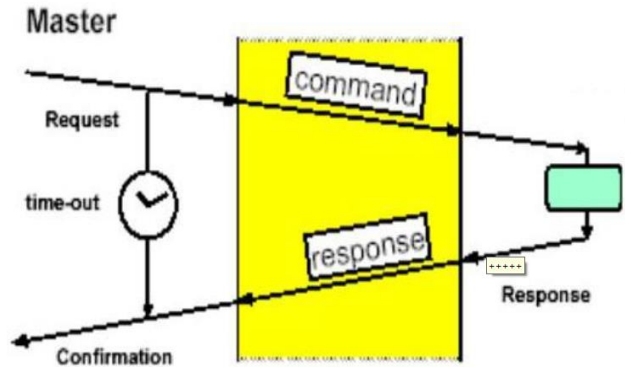

Fig.11

3.3. Transmission technologies in Profibus DP

Data transmission in this type of Profibus is performed in one of the following ways:

1. Shielded twisted-pair (STP):

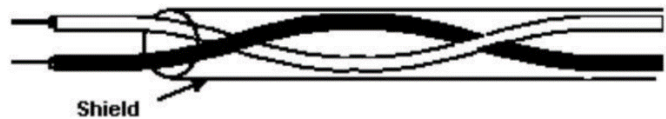

Fig.12 STP structure

2. RS485, a paired connection in the form of half-duplex, i.e., instant one, is the transmitter, and the others are receivers. Unlike RS232, this signal is a differential signal. In other words, it is not relative to the ground, but the voltage difference between the two cables, which results in lower noise acceptance.

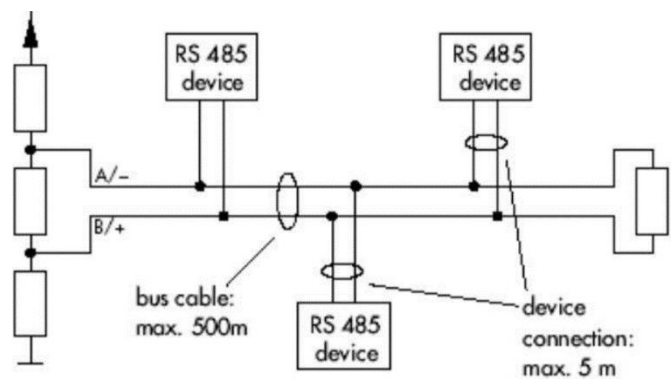

Fig.13 RS485 structure

3. Transmission using an optical fiber

## 3.4. Profibus FMS

It examines the layers 1, 2, and 7 of the ISO models. The FMS service is used in its application layer. The transmission environment and taking control of the bus are identical in FMS and DP. Thus, they can be simultaneously implemented within a network. The FMS service is used for the case where there exists a high volume of information.

### 3.4.1. Important characteristics of FMS

The transmission method is via copper cables (RS485, with the maximum speed of 1500Kbps) or optical fibers.

    1. Layers 1, 2, and 7 are used.

    2. The bus is accessed via a token pass.

    3. Includes the SDA, SDN, and SRD services.

## 3.5. The Token pass method

If we want to establish a connection between several masters, a logical (and not physical) loop is established between the network nodes. This loop's direction is determined according to node addresses and from the lowest to the highest address.

The nodes making up this loop are the masters, and the token goes from one master to another with a higher address. When the token reaches the master with the highest address, it transmits it to the master with the lowest address. Thus, a software loop is formed. In this method, each node contains a list known as LAS, which specifies the following:

- NS: The address of the next station in the token loop
- PS: The address of the previous station in the token loop
- TS: The address of the next station in the token loop

## 3.6. Profibus PA

It is an evolved example of DP that is usually used at the field level. A high level of security is naturally achieved from the IEC 1158-2 method used in this protocol because the feeding of the elements connected to this network is directly provided through a communication line.

Data transmission is performed based on the Manchester coding protocol. In this protocol, the bit 0 occurs when we have an ascending edge, while the bit 1 occurs when we have a falling edge.

In this method, the data transmission speed is constant and equal to 31.25Kbps and is not related to the cable's length. Moreover, the transmission environment may be a shielded twisted-pair (STP) cable or an unshielded UTP cable.

## 4. Conclusion

In the present review, two different types of industrial communication protocols have been proposed. There are many substantial differences between PROFIBUS and PROFINET. Currently, Profinet and Profibus are both well-proven industry leaders. Thousands of device manufacturers develop and produce products available with either a Profinet or Profibus interface. Table 1 compared different properties of these two protocols.

| Properties | Profinet | Profibus |
| --- | --- | --- |
| Physical layer | RS-485 | Ethernet |
| Speed | 12Mbit/s | 1Mbit/s or 100 Mbit/s |
| Telegram | 244 bytes | 1440 bytes |
| Address space | 126 | Unlimited |
| Technology | Master/Slave | Provider/Consumer |
| Connectivity | PA + Others | Many buses |
| Wireless | Possible | IEEE 802.11, 15.1 |
| Motion | 32 axes | >150 axes |
| Machine-to-Machine | No | Yes |

Table 1 profinet vs profibus properties

It makes sense that Profibus is more prevalent. Profinet is the younger technology, and when it was introduced, there were already tens of millions of installed Profibus devices. Between the two, Profinet is the clear winner from a technical and investment-protection perspective. Today, the adoption of Profinet is growing exponentially, and in some years, its adoption will surpass Profibus.